\documentstyle[sprocl]{article}

\def\Journal#1#2#3#4{{#1} {\bf #2}, #4 (#3)}

\def\NPB{{\em Nucl. Phys.} B}
\def\PLB{{\em Phys. Lett.}  B}
\def\PRL{\em Phys. Rev. Lett.}
\def\PRD{{\em Phys. Rev.} D}

\def\CPC{\em Comp. Phys. Comm.}
\def\be{\begin{equation}}
\def\ee{\end{equation}}
\def\bea{\begin{eqnarray}}
\def\eea{\end{eqnarray}}
\def\Dslash{\not \! \! D}

\begin{document}

\title{Fermionic Monte Carlo algorithms for lattice QCD}

\author{ Ph. de Forcrand }

\address{Swiss Center for Scientific Computing,\\
         ETH Zentrum, CH-8092 Z\"urich, Switzerland}

%%%%%%%%%%%%%%%%%%%%%%%%%%%%%%%%%%%%%%%%%%%%%%%%%%%%%%%%%%%%%%

\maketitle\abstracts{
The increase with time of computer resources devoted to simulations of
full QCD is spectacular. Yet the reduction of systematic errors is
comparatively slow. This is due to the algorithmic complexity of the
problem. I review, in elementary terms, the origin of this complexity,
and estimate it for 3 exact fermion algorithms.}

	There is a qualitative difference between quenched QCD, which
considers the dynamics of (bosonic) gauge fields, and full QCD, which
includes the dynamics of (fermionic) quarks, and thereby the physical
effects of virtual quark pair creation. The former has only short-range
interactions; the latter has long-range interactions among the gauge
fields $\{U\}$. How can this
happen, since fermionic fields $\psi$ interact with each other locally,
with action density \mbox{$\bar{\psi} (\Dslash(\{U\}) + m) \psi$ ?}
The reason is simple. 
Fermions anti-commute ($\psi(x) \psi(y) = - \psi(y) \psi(x)$),
so that they cannot be simulated directly. The standard procedure
consists in integrating them out of the partition function \\
\begin{displaymath}
\int {\cal D}\bar{\psi} {\cal D}\psi 
e^{- \sum_{x,y} \bar{\psi}(x) (\not \! D + m) \psi(y)}
\end{displaymath}
The exponential can be expanded in a power series. Because of
anti-commutation, the only surviving terms will be
$\int d\bar{\psi} d\psi \bar{\psi}(x) \psi(x) = 1$,
and the result of the integration will be 
$det (\Dslash(\{U\}) + m)$ 
for each quark flavor, or 
$det^{n_f} (\Dslash(\{U\}) + m)$
for $n_f$ flavors degenerate in mass.
This determinant cannot be factorized into local pieces, hence the
long-range interactions induced on the gauge fields.

\section{Scaling limit}\label{A}

	In the presence of fermions, QCD possesses 2 mass (or length)
scales: the string tension $\sqrt{\sigma} \sim 440 MeV$ of the pure
gauge theory, and the quark mass $m_q$. Both should be much smaller than
the inverse lattice spacing $a^{-1}$, to ensure small discretization
errors. In addition the physical box $(L a)^4$ represented by the
lattice should be large enough to contain the interesting physics.
Naively, these requirements would translate into
\be\label{req_size}
\begin{array}{cccr}
a \sqrt{\sigma} & ~~~\ll~~~ 1 ~~~\ll~~~ & L a \sqrt{\sigma} & ~~~~~~~~~~~~(a) \\
a m_q           & ~~~\ll~~~ 1 ~~~\ll~~~ & L a m_q           & ~~~~~~~~~~~~(b)
\end{array}
\ee
However because quarks are confined, the largest-size object to be
contained in the box is not a free quark, but a $q \bar{q}$ meson.
For heavy quarks, its mass is indeed $\propto m_q$. But for light
quarks, the pion mass obeys the PCAC relation
\be
m_{\pi}^2 = B m_q
\ee
So confinement helps us here, by replacing \ref{req_size}(b) by
\be\label{req_size_light}
a m_q           ~~~\ll~~~ 1 ~~~\ll~~~ L a \sqrt{B} \sqrt{m_q}
\ee
Therefore, one needs to distinguish 2 cases:
\begin{itemize}
\item heavy quarks ($m_q \gg \sqrt{\sigma}$): constraints
\ref{req_size}(a) and \ref{req_size}(b) can be summarized by
\be
a m_q           ~~~\ll~~~ 1 ~~~\ll~~~ L a \sqrt{\sigma}
\ee
\item light quarks ($m_q \ll \sqrt{\sigma}$): then \ref{req_size}(a)
and (\ref{req_size_light}) are equivalent to
\be\label{size_light}
a \sqrt{\sigma} ~~~\ll~~~ 1 ~~~\ll~~~ L a \sqrt{B} \sqrt{m_q}
\ee
\end{itemize}
We are ultimately interested in simulating the light quark case, which
corresponds to the real world, so we will try to estimate the complexity
of various fermion algorithms in that regime. Note however that
present-day simulations of full QCD are only probing the heavy-quark
domain: the crossover to light quarks, which corresponds roughly to the
strange quark mass, to the opening of $\rho \rightarrow \pi \pi$ decay,
and to dominant sea-quark effects, has not been reached yet. In any case,
the strategy for a light-quark simulation should be:\\
(1) choose the lattice spacing so that the left-hand side of inequality
(\ref{size_light}) is satisfied;\\
(2) as the quark mass is decreased, increase $L \propto m_q^{-1/2}$ to
preserve the right-hand side of the inequality.\\
I will now compare the behavior of 3 exact algorithms as 
$m_q \rightarrow 0$.

\section{The simplest algorithm: link-by-link Metropolis}\label{B}

	Consider a local change in the gauge field (say one link only).
This will induce a local change $\Delta$ in the $\Dslash$ part of the
Dirac matrix $D \equiv \Dslash + m$. The elements of matrix $\Delta$
are all zero, save for, say, $\Delta_{x_0, x_0 + \hat{\mu}}$ and
$\Delta_{x_0 + \hat{\mu}, x_0}$, keeping only spatial indices for
clarity. Consider then a Metropolis update.
The acceptance probability will be
\be
P_{acc} = min(1, \left(\frac{det(D + \Delta)}{det D}\right)^{n_f} )
\ee
So what is needed is 
\be\label{2x2}
det({\bf 1} + \Delta D^{-1}) = 
det\left( \begin{array}{cc}   
1 + \Delta_{x_0,x_0+\hat{\mu}} D^{-1}_{x_0+\hat{\mu},x_0} &
\Delta_{x_0,x_0+\hat{\mu}} D^{-1}_{x_0+\hat{\mu},x_0+\hat{\mu}} \\
\Delta_{x_0+\hat{\mu},x_0} D^{-1}_{x_0,x_0} &
1 + \Delta_{x_0+\hat{\mu},x_0} D^{-1}_{x_0,x_0+\hat{\mu}}
\end{array}\right)
\ee
Thus only 4 elements of the inverse matrix $D^{-1}$ are needed. A
straightforward way to compute them is to solve the linear systems 
\be\label{linear}
D \vec{z}_{1,2} = \vec{b}_{1,2}
\ee
where $b_{1,2}(x) = \delta_{x,x_0}$ and $\delta_{x,x_0 + \hat{\mu}}$
respectively, using an iterative solver, to sufficient accuracy.
The solution vector $\vec{z}$ represents a whole column of $D^{-1}$,
from which the useful elements can be extracted.\cite{Nakamura}

\subsection{Cost analysis}
Using this algorithm, what is the cost of generating an independent
configuration? \\
I make the plausible assumption that, as the quark mass
goes to zero, the number of iterations needed to solve (\ref{linear})
grows as $1/m_q$. This assumption certainly holds for staggered
fermions, and appears to hold also for Wilson fermions (see eg.
\cite{Melbourne}).\\
The work required per link update is then proportional to $V m_q^{-1}$,
or $V^2 m_q^{-1}$ per sweep. Since the relevant correlation length as
discussed in \ref{A} is $m_\pi^{-1} \sim m_q^{-1/2} B^{-1/2}$, the
number of sweeps necessary to decorrelate the gauge field will be 
$(m_q^{-1/2} B^{-1/2})^z$, or $\sim m_q^{-1}$, since $z \approx 2$ for a
Metropolis algorithm. Altogether, the work per independent configuration
is
\be
	V^2 m_q^{-2}
\ee
or, since the lattice size $L$ must scale as $m_q^{-1/2}$
(see eq.\ref{size_light})\\
\be\label{compl_1}
	L^{12}
\ee
or $a^{-12}$.
\footnote{For a non-confining theory in $d$ dimensions, 
the correlation length would
become $m_q^{-1}$, giving a cost per independent configuration 
$V^2 m_q^{-3}$; the lattice size would scale as $m_q^{-1}$, giving a
complexity $L^{2d + 3}$.}
The prefactor can be considerably reduced in clever variants of this 
method.\cite{CPC} But the algorithmic complexity remains daunting.
It is even possible that (\ref{compl_1}) actually underestimates it.
The statement that the number of sweeps per
independent configuration grows as $m_q^{-1}$ relies on the assumption
that the change in each link update is not limited by further
constraints. On the other hand a naive inspection of the determinant
(\ref{2x2}) indicates that $\Delta$ might scale as $m_q$, to compensate
the $m_q^{-1}$ divergence of $D^{-1}$. If the step size is $\sim m_q$,
the number of steps needed to explore the whole gauge group and obtain a
decorrelated configuration would be $\sim m_q^{-2}$, instead of
$m_q^{-1}$ as stated above. In that case the complexity would be\\
\be
	V^2 m_q^{-3} ~~~{\rm or}~~~ L^{14}
\ee

Whether in practice the step size goes like $\sqrt{m_q}$ or $m_q$, it is
interesting that this algorithm, which in principle admits Monte Carlo
steps of any size, automatically restricts them as the quark mass
decreases. This restriction creates further problems, of ergodicity and
of slowing-down, in the presence of an energy barrier like a zero-mode
of the Dirac matrix separating successive topological sectors. 
These problems are entirely neglected here: a comparative study of 
fermionic algorithms with regard to exploring topological sectors is
still awaiting.

\section{Hybrid Monte Carlo}\label{C}

	The magic of Hybrid Monte Carlo (HMC\cite{HMC}) consists in updating all
links simultaneously by a very small step, which allows a simplifying
linearization of the problem. The error caused by the finiteness of the
step size is periodically corrected by a Metropolis test.
HMC is the standard fermionic algorithm. It has been reviewed several
times, and its complexity conjectured\cite{Creutz,Rajan} and studied
numerically.\cite{Gupta} 

HMC first introduces one species of auxiliary bosonic fields $\phi$,
through the Gaussian integral\\
\be
det^2(\Dslash + m) = constant \int {\cal D}\phi^\dagger {\cal D}\phi
e^{- \phi^\dagger [(\Dslash+m)^\dagger (\Dslash+m)]^{-1} \phi}
\ee
The squaring of $(\Dslash + m)$ guarantees the convergence of the
Gaussian integral (since it may happen, for small quark masses, that the
lattice Dirac operator develop eigenvalues with a negative real part).
It also makes it very cheap to update the field $\phi$ by a heatbath:
$\phi \leftarrow (\Dslash+m)^\dagger \eta$, where $\eta$ is a complex
Gaussian vector.

Heatbath updates of $\phi$ will alternate with updates of the gauge
links. This is accomplished by introducing fictitious momenta
$p_{x,\mu}$ conjugate to the gauge fields $A_{x,\mu}$ 
($U_{x,\mu} = exp(i A_{x,\mu})$), and a fictitious Hamiltonian\\
\be
H = \sum_{x,\hat{\mu}} \frac{p_{x,\hat{\mu}}^2}{2}
+ S_G + \phi^\dagger [(\Dslash+m)^\dagger (\Dslash+m)]^{-1} \phi
\ee
where $S_G$ is the gauge action. The last 2 terms form the potential
energy ${\cal V}$ of the gauge fields. After a heatbath update of
$\phi$, the momenta $p_{x,\mu}$ are initialized, also by heatbath
$p_{\mu} \leftarrow \eta$. Then a discrete integration of Hamiltonian
evolution is pursued along a ``trajectory" of length ${\cal O}(1)$ in
fictitious time, using generally the leapfrog integrator:
\bea
	A(t + \frac{\delta\tau}{2}) & = & A(t) + \frac{\delta\tau}{2} p \\
	p(t + \delta\tau) & = & p(t) - \delta\tau \nabla {\cal V} \\
	A(t + \delta\tau) & = & A(t + \frac{\delta\tau}{2}) 
+ \frac{\delta\tau}{2} p 
\eea
The algorithm, at this point, is exact up to integration errors,
which cause in particular violations $\Delta E$ of the conservation of
the total Hamiltonian energy. These deviations are compensated for by a
Metropolis test at the end of the trajectory, with acceptance
$P_{acc} = min(1, e^{-\Delta E})$. Thus HMC can be viewed as an
elaborate Metropolis scheme, where the candidate configuration $\{U_{new}\}$
is obtained from $\{U_{old}\}$ not by a random step, but by a carefully
guided step. Detailed balance is satisfied because the leapfrog
integrator is symplectic and reversible, guaranteeing the ``evenness" of
the distribution of proposed changes.

\subsection{Cost analysis}
\begin{itemize}
\item The optimal step size is the result of a compromise between fast
Hamiltonian integration and high Metropolis acceptance. Very briefly,
the RMS energy violation per step is 
$\delta E \sim \sqrt{V} m_q^{-3} \delta \tau^3$.
The $\sqrt{V}$ comes from the uncorrelated contributions over a large
volume. The $m_q^{-3} \delta \tau^3$ comes from the leading error in a
reversible (second-order) integrator. Over a trajectory of $1/\delta
\tau$ steps, the RMS energy violation is then 
$\Delta E \sim \sqrt{V} m_q^{-3} \delta \tau^2$,
which must be kept constant to preserve a constant acceptance.
Hence $\delta \tau \sim V^{-1/4} m_q^{3/2}$.\cite{Creutz,Rajan,Gupta}
\item The cost per step comes overwhelmingly from the calculation of the
fermionic force, requiring ${\cal O}(m_q^{-1})$ iterations of a linear
solver.
\end{itemize}
The total cost per trajectory of length 1 is thus 
$\sim V^{5/4} m_q^{-5/2}$.\\
Different scenarios have been proposed as to the number of trajectories
necessary for decorrelation. \\
(a) Trajectories of length ${\cal O}(1)$ are sufficient to guarantee a
dynamical critical exponent $z = 1$. The correlation length is 
$\sim m_q^{-1/2}$, so that the total cost is 
\be
	V^{5/4} m_q^{-3} ~~~{\rm or}~~~ L^{11}
\ee
since $m_q^{-1/2} \propto L$.\\
(b) If one increases the trajectory length like the correlation length,
then $z = 0$.\cite{Duane} The step size becomes 
$\delta \tau \sim V^{-1/4} m_q^{7/4}$, and the cost is
\be
	V^{5/4} m_q^{-13/4} ~~~{\rm or}~~~ L^{11.5}
\ee
(c) With trajectory of length ${\cal O}(1)$, the critical exponent $z$ is
2. Then the cost is
\be
	V^{5/4} m_q^{-7/2} ~~~{\rm or}~~~ L^{12}
\ee
In my opinion, scenario (a) is too optimistic (a similar scenario for
the quenched theory has been numerically proven wrong\cite{TARO}).
I favor (c).

HMC is straightforward to program, even with improved, less local gauge
or fermionic actions. It directly benefits from recent progress in
linear solvers. It can be further accelerated, by trading the Dirac
matrix $D$ for another one of same determinant, but giving a smaller
force. The replacement of $D$ by an $L D U$ combination allows a large
(${\cal O}(5)$) increase of the step size.\cite{Takaishi}\\
One potential problem with HMC on large lattices comes from round-off
errors. The Metropolis acceptance depends on the energy difference
$\Delta E$, which is a small ${\cal O}(1)$ difference of 2 large numbers
${\cal O}(V)$. Special care must be taken to calculate $\Delta E$
accurately enough that violations of reversibility remain 
negligible.\cite{SESAM,Kennedy} The next algorithm circumvents this problem.

\section{The multiboson approach}\label{D}

	This approach, originally proposed by L\"uscher\cite{Luscher},
can be summarized as follows (see \cite{Galli} for a detailed
presentation):
\begin{itemize}
\item Choose a Chebyshev-like polynomial $P_n(x)$ approximating $1/x$ in
a domain of the complex plane which includes the eigenvalue spectrum of
the Dirac operator $D$. $P_n$ can be factorized as
\be
	P_n(x) = ~constant~ \Pi_{k=1}^n (x - z_k)
\ee
\item It follows, by going to an eigenbasis of $D$, that
\be
det D \approx ~constant~ \Pi_{k=1}^n det^{-1} (D - z_k {\bf 1})
\ee
or
\be
|det D|^2 \approx ~constant~ \Pi_{k=1}^n 
det^{-1} (D - z_k {\bf 1})^\dagger (D - z_k {\bf 1})
\ee
Each factor in the right-hand side can be replaced by a Gaussian
integral over an auxiliary field $\phi_k$. The effective action is then
\be
S_{eff} = \beta S_G + \sum_{k=1}^n 
\phi_k^\dagger (D - z_k {\bf 1})^\dagger (D - z_k {\bf 1}) \phi_k
\ee
\item Update the $U$'s and the $\phi_k$'s by a local MC algorithm
(typically a mixture of over-relaxation and heatbath) for a reversible
sequence of steps (``trajectory").
\item Correct the error $|det D P_n(D)|^2$ in the measure by a cheap
Metropolis test at the end of each trajectory.
\end{itemize}

In addition to avoiding the accuracy problems of HMC, this method relies
on manifestly finite update steps, so there is hope that it will be more
efficient in overcoming energy barriers. Moreover, it generalizes
readily to other functions of $det D$, as needed for simulations of odd
numbers of flavors\cite{nf=1} or SUSY theories.\cite{Montvay} The
price to pay for these potential advantages is a large increase in
memory (the number of auxiliary fields needed is about the same as the
number of iterations of the linear solver in HMC), and a programming
complexity which increases very rapidly with the range of the operator
$D$.

\subsection{Cost analysis}
\begin{itemize}
\item The work per step grows like $n$, the number of bosonic fields.
\item During a link update, the bosonic fields are kept frozen. They
limit the size of the update step to $\propto n^{-1/2}$, so that the
autocorrelation time of the $U$'s grows linearly with $n$.\cite{Jeger,Galli}
\item The optimal choice for $n$ is therefore a compromise between fast
evolution and low Metropolis acceptance. The approximation error
$|\lambda P_n(\lambda) - 1|$ for the Chebyshev-like polynomials
considered decreases exponentially with $n$, so that it will be bounded
by ${\cal O}(e^{-c n m_q})$ for some constant $c$. To keep the
Metropolis acceptance constant, one must preserve
$Log( det D P_n(D) ) \sim V e^{-c n m_q}$. Therefore, 
$n \propto m_q^{-1} Log V$.
\item One expects additional slowing down $m_q^{-z}$ from the local MC
dynamics of fields $\phi_k$. 
\end{itemize}
Altogether these factors yield a cost per independent configuration
$V (Log V)^2 m_q^{-2-z}$. Since the $\phi_k$ dynamics are local, one can
expect at best $z = 1$, which is consistent with numerical 
observations.\cite{Galli}  A more conservative scenario would be $z = 2$. 
Clearly, a cluster algorithm with $z < 1$ would make this algorithm very
competitive. In any case the cost is, for $z = 1$
\be
	V (Log V)^2 m_q^{-3} ~~~{\rm or}~~~ L^{10} (Log L)^2
\ee
or, for $z = 2$
\be
	V (Log V)^2 m_q^{-4} ~~~{\rm or}~~~ L^{12} (Log L)^2
\ee

\section{Summary}\label{E}

Table 1 summarizes our analysis. It is remarkable that all
algorithms ultimately give an $L^{12}$ dependence as the quark mass
approaches 0. The reason for this probably is that all 3 algorithms are
equally incapable of accelerating the evolution of the relevant
approximate zero-modes of the Dirac operator. The implementation of the
Cornell program\cite{Cornell} (Fourier acceleration of MC dynamics, after
Fourier-accelerated gauge-fixing) could improve this picture. On the
other hand, the relevant dynamics as $m_q \rightarrow 0$ is most likely
the motion through topological sectors, which our analysis ignores.

Topological sectors are separated by narrow but high energy barriers.
The dynamics of barrier-crossing is likely to be sensitive to the size
of an elementary update step, which is listed in Table 2.
Lack of ergodicity appears especially dangerous with HMC, as noted some
time ago.\cite{Teper}

\begin{table}[tbh]\label{tabl1}
\caption{Complexity of fermionic algorithms - conservative scenario.}
\vspace{0.4cm}
\begin{center}
\begin{tabular}{|c|c|c|c|}
\hline
& Link-by-link & HMC & Multiboson \\
 \hline
$m_q$ fixed & $V^2 m_q^{-2}$ & $V^{5/4} m_q^{-7/2}$ & $V (Log V)^2 m_q^{-4}$ \\
 \hline
$L \propto m_q^{-1/2}$ & $L^{12}$ & $L^{12}$ & $L^{12}$ \\
 \hline
\end{tabular}
\end{center}
\end{table}

\begin{table}[tbh]\label{tabl2}
\caption{Size of an elementary update step.}
\vspace{0.4cm}
\begin{center}
\begin{tabular}{|c|c|c|c|}
\hline
& Link-by-link & HMC & Multiboson \\
 \hline
$m_q$ fixed & $m_q^{1/2}$ & $V^{-1/4} m_q^{3/2}$ & $(Log V)^{-1/2} m_q^{1/2}$ \\
 \hline
$L \propto m_q^{-1/2}$ & $L^{-1}$ & $L^{-2}$ & $L^{-1} (Log V)^{-1/2}$ \\
 \hline
\end{tabular}
\end{center}
\end{table}

For current simulations, with not-so-light quarks in not-so-large volumes, 
the first line of Table 1, which separates the
complexity in volume and in quark mass, may be the more relevant. The 3
algorithms reviewed, presented in chronological order, progressively
trade an increased dependence on the quark mass for a decreased
dependence on the volume. This represents progress for fixed quark mass
and large volumes (meaning $m_\pi L a \gg 1$): in that case, the
multiboson method should be the most efficient. \\
This has in fact been verified in the current study of 1-flavor QCD
thermodynamics, with heavy dynamical quarks.\cite{nf=1}  The dynamics of
the Polyakov loop were $\sim 10$ times faster with the multiboson method
than with HMC. \\
On the other hand, for very light quarks (in a small volume), HMC and
the multiboson methods perform equivalently.\cite{Galli-Jansen}

Finally, I stress that these are costs per independent
configuration. The number of independent configurations required to
preserve the statistical signal-to-noise ratio may increase as $m_q$
decreases, depending on the observable.

\section{Conclusion}

	Lattice QCD simulations started in 1980, at the initiative of
one of this meeting's participants.\cite{Creutz1980}
Progress in algorithms has since been as impressive as in hardware.\\
On one hand, lattice QCD is a homogeneous problem on a regular grid: it
belongs to the class of ``embarrassingly parallel" problems, for which
the simplest, cheapest SIMD parallel computer is sufficient. Current
simulation projects call for ${\cal O}$(1 Gigaword) of data, ${\cal O}$(a
few months) of computer time on ${\cal O}$(a few 100) GFlops machine. The
resources involved justify continued research on algorithms.\\
On the other hand, the current situation is quite different in the
quenched and the full QCD cases. Quenched QCD is a ``mature" field:
algorithms are stagnant; a continuum extrapolation of MC results, in
large volumes, is reliable at the $10\%$ level. Full QCD still is at an
early stage: algorithms are still being explored and improved; 8 years
elapsed before a practical exact algorithm (HMC) was devised; 8 more
years passed by before a competitive alternative (multiboson) was found;
more progress will undoubtedly come. The cost of current algorithms is
still so high that simulations are generally squeezed in a dilemma:
either the lattice is too coarse, or the physical volume is too small.
The interesting regime of light dynamical quarks is barely starting to
be explored. This slower progress is a consequence of the tremendous complexity
$\sim L^{12}$ of full QCD (a similar analysis for quenched QCD yields
a complexity $\sim L^6$ ``only"). 

The relevance of the complexity analysis presented in this review is
limited. Table 1 only shows asymptotic behaviors (large volume,
small quark mass). In any case, prefactors must be determined through
numerical experiment.\cite{Melbourne,Jansen} But one outcome should be
clear: an improved action, which would reduce discretization errors such
that the lattice spacing can be doubled, offers a potential savings of
$2^{12} \sim 4000$ in computer time.

\section*{Acknowledgments}
I am grateful to ZiF and to the Physics Dept. of the University of
Bielefeld for their kind hospitality while this paper was finally
written. I thank Atsushi Nakamura for comments.

\end{document}